\documentclass[12pt]{article}
\usepackage[centertags]{amsmath}
\usepackage{amsfonts}
\usepackage{amssymb}
\usepackage{amsthm}
\newcommand{\beq}{\begin{equation}}
\newcommand{\eeq}[1]{\label{#1}\end{equation}}
\newcommand{\bea}{\begin{eqnarray}}
\newcommand{\eea}[1]{\label{#1}\end{eqnarray}}
\begin{document}
\setlength{\topmargin}{-1cm} \setlength{\oddsidemargin}{0cm}
\setlength{\evensidemargin}{0cm}
\begin{titlepage}
%\hfill  {\small NYU-TH/07-137}
\begin{center}
{\Large \bf Intrinsic Cutoff and Acausality for Massive Spin 2
Fields Coupled to Electromagnetism}

\vspace{20pt}

{\large Massimo Porrati and Rakibur Rahman}

\vspace{12pt}

Center for Cosmology and Particle Physics\\
Department of Physics\\ New York University\\
4 Washington Place\\ New York, NY 10003, USA

\end{center}

\vspace{20pt}

\begin{abstract}
We couple a massive spin 2 particle to electromagnetism. By
introducing new, redundant degrees of freedom using the
St\"uckelberg formalism, we extract an intrinsic, model independent
UV cutoff of the effective field theory describing this system. The
cutoff signals both the onset of a strongly interacting  dynamical
regime and a finite size for the spin 2 particle. We show that the
existence of a cutoff is strictly connected to other pathologies of
interacting high-spin fields, such as the Velo-Zwanziger acausality.
We also briefly comment on implications of this result for the
detection of high spin states and on its possible generalization to
arbitrary spin.
\end{abstract}

\end{titlepage}

\newpage
\section{Introduction}
While the coupling of massless high-spin particles to
electromagnetism or gravity is notoriously fraught with
inconsistencies (see e.g.~\cite{ad}), massive charged particles of
any spin can and do exist. Massive higher-spin particles like
$\pi_2$(1670), $\rho_3$(1690) or $a_4$(2040) have amply been produced
in particle colliders. They are resonances, thus composite and
unstable. So they can be described by an effective, local field
theory only up to some finite UV cutoff $\Lambda$, of the order of
their inverse size. In known resonances, this cutoff is also of the
same order of magnitude as their mass $m$.

String theory also predicts massive higher-spin particles, with mass
at least as large as $\mathcal{O}(M_{string})$. These particles
couple to $U(1)$ gauge fields or gravity, and  can be given an
effective field theory description, but again with a finite cutoff
$\Lambda=\mathcal{O}(M_{string})$.

Both in string theory and in QCD, high spin states always interact
with other states of lower spin and mass $\mathcal{O}(\Lambda)$. We
may wonder if the approximate equality $\Lambda\approx m$ is just a
property of these two examples, or if it is a general feature of
charged high-spin particles. An answer to this question is relevant
to figuring out possible experimental signatures of high-spin
particles in future colliders. For instance, it rules out long-lived
high-spin charged particles, and thus affects directly the strategy
for their search.

In this paper we will begin a study of interacting high-spin massive
particles, starting with a relatively simple yet interesting case: a
spin two particle coupled to electromagnetism. The first task in
constructing an effective field theory for a high-spin field is to
write a free Lagrangian. This is known for massive fields of
arbitrary spin. The earliest such Lagrangian was written a long time
ago by Singh and Hagen~\cite{sh}.
 Auxiliary fields are necessary to the best of our
knowledge, unless the Lagrangian is nonlocal~\cite{f}.
Gauge-invariant Lagrangians for free {\em massless} high-spin fields
are also well-known~\cite{ff}. Inconsistencies arise when one tries to
make these fields interact. These inconsistencies are due to the
absence of currents invariant under the high-spin gauge symmetry, as
first clearly illustrated in the case of spin 5/2 coupled to gravity
by Aragone and Deser in~\cite{ad}\footnote{The absence of gauge
invariant currents also makes the coupling of massless $s=3/2,2$
fields to electromagnetism inconsistent.}.

Massive high-spin fields {\em can} be coupled to electromagnetism.
After all, charged high-spin resonances do exist! This fact alone
implies that any Lagrangian describing charged high-spin fields
interacting with electromagnetism must be singular in the massless
limit $m\rightarrow 0$. If one takes the Singh-Hagen Lagrangians and
follows the minimal-coupling prescription to introduce
electromagnetic interactions, then the massless singularity is far
from manifest. Indeed, the resulting Lagrangians contain only
positive powers of the mass. Scattering amplitudes become singular
because in the massless limit the high-spin free Lagrangian acquires
a gauge invariance that makes its kinetic term non-invertible.
Correspondingly, the propagator of the massive theory also becomes
singular when $m\rightarrow 0$. The problem is thus associated with
the gauge invariance of the free theory. It thus manifests already
for massive spin one; but in that case it can be cured by adding a
non-minimal dipole term. For higher spins, some but not all singular
terms can also be eliminated by adding non-minimal terms~\cite{fpt}.

In this paper, we will argue that the massless singularity is indeed
robust and cannot be completely canceled by adding non-minimal
terms. We will also quantify the degree of singularity of the
massless limit. Specifically, for the case of spin 2, we will argue
that the cutoff of the effective action is always lower than \beq
\Lambda_2 \equiv \frac{m}{\sqrt{e}}, \eeq{m1} where $e$ is the
electric charge. We will always work in flat space; equivalently, we
will consider particles with mass much higher than the inverse
curvature radius of the space-time background. For different values
of the cosmological constant additional consistency bounds also
apply~\cite{d2}.

A systematic study of the mass singularity is
greatly facilitated by using the St\"uckelberg formalism, i.e., by
making the massive free theory gauge invariant through the addition
of auxiliary fields. These fields can be set to vanish using the
resulting gauge invariance. In this case the St\"uckelberg action
reduces to the original one. On the other hand, a different,
judicious choice of (covariant) gauge fixing can make all kinetic
terms in the theory canonical  (diagonal on momentum eigenstates and
proportional to $p^\mu p_\mu$). In this case, inverse powers of the
mass appear explicitly in the (non-renormalizable) interaction terms
involving the auxiliary fields.

A simple, standard example of the procedure is a complex, massive
spin 1 field $W_\mu$ coupled to electromagnetism. The free
Lagrangian is \beq L = -\frac{1}{2}G_{\mu\nu}^* G^{\mu\nu} - m^2
W_\mu^* W^\nu, \qquad G_{\mu\nu}=\partial_\mu W_\nu -\partial_\nu
W_\mu. \eeq{m2}

The action becomes gauge invariant after a complex compensator
scalar field $\phi$ is introduced by the substitution $W_\mu =V_\mu
-\partial_\mu \phi/m$. After adding the gauge fixing term
$-|\partial_\mu V^\mu - m\phi|^2$ the action eq.~(\ref{m2}) becomes
diagonal \beq L- |\partial_\mu V^\mu - m\phi|^2 = V_\mu^* (\Box -
m^2)V^\mu + \phi^* (\Box - m^2) \phi. \eeq{m3} The minimal
substitution $\partial_\mu \rightarrow D_\mu \equiv \partial_\mu \pm
ieA_\mu$ generates non-renormalizable interaction
terms\footnote{They are non-renormalizable because the fields
$V_\mu,\phi$ have canonical dimension one, and their kinetic terms
are nonsingular and canonically normalized.}: \beq \left[
-i\frac{e}{2m} F_{\mu\nu}\phi^* (D^\mu V^\nu - D^\nu V^\mu) + c.c.
\right] - \frac{e^2}{2m^2}F_{\mu\nu}F^{\mu\nu}\phi^* \phi. \eeq{m4}
Notice that these terms arise only from the kinetic term of
Lagrangian~(\ref{m2}), {\em not} from the gauge fixing.

The UV cutoff signaling the breakdown of our effective field theory
is now explicit: it is the coupling constant multiplying the
non-renormalizable interactions terms. To make this even clearer, we
take the decoupling limit \beq m\rightarrow 0,\qquad
e\rightarrow 0, \qquad \frac{m}{e} = \mbox{constant}\equiv \Lambda.
\eeq{m5} The Lagrangian does not become free, but instead reduces to
\beq L=V_\mu^* \Box V^\mu + \phi^* \Box \phi + \left[
-\frac{i}{2\Lambda} F_{\mu\nu}\phi^* (\partial^\mu V^\nu -
\partial^\nu V^\mu) + c.c. \right] -
\frac{1}{2\Lambda^2}F_{\mu\nu}F^{\mu\nu}\phi^* \phi. \eeq{m6}

For spin one the cutoff $\Lambda$ is not an intrinsic property of
the theory, since all non-renormalizable terms are canceled by
adding to the Lagrangian a non-minimal (dipole) term \beq
L_{NM}=ieF_{\mu\nu} W^{*\mu} W^\nu. \eeq{m7} This term is power
counting renormalizable; so it does not introduce new massless
divergences. By contrast, a quadrupole term $\sim (e/m)\partial F
W^* \partial W$ gives rise to the singular term $\sim (e/m)\partial
F V^* \partial V$, i.e., it introduces new non-renormalizable
interactions, some of which do not involve the St\"uckelberg scalar
$\phi$.

Instructed by this example, we can lay down a general procedure for
studying the dynamics of charged high-spin fields. The same
procedure will also apply to other interactions of high-spin fields
(gravitational interactions, for instance\footnote{Massive gravity,
that is a self-interacting massive spin 2, was studied using a
St\"uckelberg-like formalism in~\cite{ahgs}.}). Modulo technical
complications due to the presence of auxiliary fields, it is
essentially what we followed in the spin 1 case.

\subsubsection*{A Seven-Step Prescription}
\textbf{Step 1}: Write a (non-gauge invariant) massive Lagrangian
with minimal number of auxiliary fields (e.g., {\em \`a la}
Singh and Hagen).
\\
\textbf{Step 2}: Introduce St\"uckelberg fields and St\"uckelberg
gauge symmetry. Any auxiliary field (appearing in the Lagrangian in
step 1) that is not a trace of the high-spin field must be
identified as a trace of a St\"uckelberg field. For such a field one
obtains a gauge
invariance for free~\cite{pr}.\\
\textbf{Step 3}: Complexify the fields, if required, and introduce
interaction with a new gauge field (e.g., electromagnetism) by
replacing ordinary
derivatives by their covariant counterparts.\\
\textbf{Step 4}: Diagonalize all kinetic terms, i.e., get rid of
kinetic
mixing by field redefinitions and/or by gauge-fixing terms.\\
\textbf{Step 5}: Look for the most divergent term(s) in the
Lagrangian, in an appropriate limit of zero mass and zero coupling.
These terms will involve fields that are zero (i.e. gauge) modes of
the free kinetic operator before gauge fixing. One needs to take
care of the non-commutativity of covariant derivatives and correctly
interpret terms
proportional to the equations of motion.\\
\textbf{Step 6}: Try to remove non-renormalizable terms by adding
non-minimal terms. This
may not always be possible.\\
\textbf{Step 7}: Find the cutoff of the effective field theory, and
interpret the physics implied by the divergent term(s).

The procedure just outlined above also summarizes our paper.
Specifically, steps 1 through 5 will be carried over for the case of
a charged spin 2 field in Section 2. Section 3 will carry over steps
6 and 7, while the concluding Section 5 will add some more comments
on the physics of interacting high-spin fields. Section 4 is somehow
at variance with the rest of the paper: it shows that another
well-known problem of high-spin fields interacting with
electromagnetism, namely the Velo-Zwanziger acausality~\cite{vz,v},
can also be addressed by our formalism.

\section{Massive Spin 2 Field Coupled to Electromagnetism}
The electromagnetic coupling of a charged massive spin 2 field has
been studied in ~\cite{vz,v,fe,ks,sc}. Such theories are unavoidably
fraught with difficulties: Velo and Zwanziger~\cite{vz,v}, for
example, concluded that in a fixed, external electromagnetic
background charged massive spin 2 particles show pathological
behavior like superluminality and/or acausality. However, as one
employs the St\"{u}ckelberg description, the same underlying physics
has a new interpretation. In particular, now the theory can be
trusted only up to some intrinsic cutoff, that cannot be sent to
infinity. The ``pathologies'' arise when one (mistakenly) tries to
extrapolate an effective description based on a local Lagrangian
beyond its regime of validity. In this section and the next we are
going to investigate the physics of massive spin 2 field coupled to
electromagnetism using this new formalism.

Our starting point is the Pauli-Fierz Lagrangian~\cite{pf, n}, which
is the unique ghost-free, tachyon-free Lagrangian for massive spin 2
field. \beq L= -\frac{1}{2} (\partial_\mu
h_{\nu\rho})^2+(\partial_\mu h^{\mu\nu})^2 +\frac{1}{2}(\partial_\mu
h)^2\, -
\partial_\mu h^{\mu\nu}\partial_\nu h
-\frac{m^2}{2}[h_{\mu\nu}^2-h^2], \eeq{r1} where $h=h^\mu_\mu$. This
Lagrangian does not have any manifest gauge invariance. Now by the
field redefinition \beq h_{\mu\nu} \rightarrow
\tilde{h}_{\mu\nu}=h_{\mu\nu} +
\frac{1}{m}\,\partial_{\mu}\left(B_\nu-\frac{1}{2m}\partial_{\nu}\phi\right)
+
\frac{1}{m}\,\partial_{\nu}\left(B_\mu-\frac{1}{2m}\partial_{\mu}\phi\right),
\eeq{r2} we create a gauge invariance \bea \delta h_{\mu\nu} &=&
\partial_{\mu}\lambda_{\nu}+\partial_{\nu}\lambda_{\mu},\label{r3} \\
\delta B_{\mu} &=& \partial_{\mu}\lambda - m\lambda_{\mu},
\label{r4}
\\
\delta\phi &=& 2m\lambda . \eea{r5}

This gauge invariance, which we will refer to as the St\"{u}ckelberg
symmetry, has been obtained by introducing new (St\"{u}ckelberg)
fields $B_{\mu}$ and $\phi$, which can always be gauged away. Yet,
as pointed out in the introduction, introducing this redundancy will
allow us to unveil the dangerous degrees of freedom and interactions
hidden inside the spin 2 action.

It is worth pointing out that one can obtain the St\"{u}ckelberg
version of the Pauli-Fierz Lagrangian by simply starting from a
linearized Einstein-Hilbert action in (4+1)D, and then Kaluza-Klein
reduce it to (3+1)D~\cite{ady, rs1}. All the fields $h_{\mu\nu}$,
$B_{\mu}$ and $\phi$ thus come from a single higher dimensional
massless spin 2 field. The higher dimensional gauge invariance
translates itself into the St\"{u}ckelberg symmetry eqs.~(\ref{r3},
\ref{r4}, \ref{r5}) in lower dimension. This observation is not
trivial; in particular, in the case of higher spins~\cite{pr}, this
may help us to construct consistent Lagrangians that can be readily
coupled to a $U(1)$ field or gravity, while maintaining at the same
time the covariant version of the St\"{u}ckelberg symmetry.

To couple our massive spin 2 field to electromagnetism we first
complexify the Pauli-Fierz Lagrangian with St\"{u}ckelber fields,
and then replace ordinary derivatives by covariant ones. \beq
\partial_{\mu} \rightarrow D_{\mu}
\equiv \partial_{\mu} \pm ieA_\mu . \eeq{r6} We have \beq L=
-\,|D_\mu \tilde{h}_{\nu\rho}|^2+2|D_\mu \tilde{h}^{\mu\nu}|^2
+|D_\mu \tilde{h}|^2 - [D_\mu \tilde{h}^{*\mu\nu}D_\nu \tilde{h} +
c.c.] -
m^2[\tilde{h}_{\mu\nu}^*\tilde{h}^{\mu\nu}-\tilde{h}^*\tilde{h}]
-\frac{1}{4}F_{\mu\nu}^2,
\eeq{r7} with \beq \tilde{h}_{\mu\nu} = h_{\mu\nu} +
\frac{1}{m}\,D_{\mu}\left(B_\nu-\frac{1}{2m}D_{\nu}\phi\right) +
\frac{1}{m}\,D_{\nu}\left(B_\mu-\frac{1}{2m}D_{\mu}\phi\right).
\eeq{r8} Lagrangian~(\ref{r7}) now enjoys a covariant
St\"{u}ckelberg symmetry:
\bea
\delta h_{\mu\nu} &=&
D_{\mu}\lambda_{\nu}+D_{\nu}\lambda_{\mu},\label{r9} \\
\delta B_{\mu} &=& D_{\mu}\lambda - m\lambda_{\mu}, \label{r10} \\
\delta\phi &=& 2m\lambda. \eea{r11}

The above symmetry is obvious only because we had at hand a
convenient form of the St\"{u}ckelberg Lagrangian. This is not so
plain in the case of spin 3 and higher, where auxiliary fields,
which are not traces of the high-spin field, are unavoidable. At
this point it is important to note that the authors in Ref.~\cite{z}
also considered the gauge invariant description to investigate
consistent theories of interactions of massive high spin fields.
Starting with a St\"{u}ckelberg invariant \emph{free} theory, they
used minimal substitution: $\partial_{\mu} \rightarrow D_{\mu}$, to
couple the theory to a gauge field. But as the resulting Lagrangian
does \emph{not} have St\"{u}ckelberg invariance, one has to look for
new terms, which must be added to the Lagrangian to recover the
invariance. On the other hand, our approach by construction
guarantees that St\"{u}ckelberg symmetry is intact by the minimal
substitution. While this may not seem to be a significant
achievement for the simple case of spin 2, as one considers higher
spins~\cite{pr} the elegance of our method tremendously facilitates
the job of writing down a St\"{u}ckelberg invariant interacting
Lagrangian. In any case, our goal is not just to obtain a gauge
invariant description, but to employ the latter to extract a model
independent UV cutoff of the effective field theory describing the
high spin system, and to show how the well-known pathologies are
related to the very existence of a cutoff.

We will now explicitly work out the various terms in the Lagrangian.
In doing so we keep in mind that covariant derivatives do not
commute. \beq [D_{\mu} , D_{\nu}]=\pm ieF_{\mu \nu}. \eeq{r12} This
of course introduces an ambiguity in the definition of the minimal
Lagrangian~(\ref{r7}). More generally, the Lagrangian is ambiguous
because one can always add to it terms vanishing at $F_{\mu\nu}=0$.
We will exploit this ambiguity in Section 3.

After a few integrations by parts we arrive at \bea L &=&
-|\partial_\mu h_{\nu\rho}|^2 + 2|\partial_\mu h^{\mu\nu}|^2 +
|\partial_\mu h|^2 - [\partial_\mu h^{*\mu\nu}\partial_\nu h +
c.c.] - m^2[h_{\mu\nu}^*h^{\mu\nu}-h^*h] - \frac{1}{4}F_{\mu\nu}^2 \nonumber \\
& & -2(|\partial_\mu B_\nu|^2-|\partial_\mu B^\mu|^2) +
[\phi^*(\partial_\mu\partial_\nu h^{\mu\nu}-\Box h) + c.c.] \nonumber \\
& & +2m[B_{\mu}^*(\partial_\nu h^{\mu\nu}-\partial^\mu h) +
c.c.]+L_{int}. \eea{r13} Here $L_{int}$ is the interaction
Lagrangian. It consists of various terms, each one containing at
least one power of $e$, and possibly an inverse power of
$m\,$($1/m^4$ at most). These are the terms we are interested in.
Before we write down the interaction terms explicitly, let us
concentrate on the kinetic terms.

It is important that the kinetic terms be diagonalized. This makes
sure that the propagators in the theory have good high energy
behavior\footnote{I.e., that all propagators are proportional to
$1/p^2$ for momenta $p^2 \gg m^2$.}. Then we can assign canonical
dimensions to  the higher-order operators in the interaction
Lagrangian, so that we can interpret ours as an effective field
theory valid up to some cutoff determined by the most divergent terms
in the $m\rightarrow 0$ limit. The kinetic mixings between $\phi$
and $h_{\mu\nu}, h$ can be eliminated by a standard field
redefinition \beq h_{\mu\nu} \rightarrow h_{\mu\nu} -
\frac{1}{D-2}\,\eta_{\mu\nu}\phi, \eeq{r14} where $D=4$ is the
space-time dimensionality. This  also generates a kinetic term for
$\phi$ with the correct sign. The free part of the Lagrangian now
looks like \bea L_{free} &=&L_{PF} - \frac{1}{4}F_{\mu\nu}^2 -
2(|\partial_\mu B_\nu|^2-|\partial_\mu B^\mu|^2) +
2m\left[B_{\mu}^*\left(\partial_\nu h^{\mu\nu}-\partial^\mu h +
\frac{3}{2}\partial_\mu\phi\right) + c.c.\right]\nonumber \\
&&-\,\frac{3}{2}|\partial_\mu\phi|^2 - \frac{3}{2}m^2[\phi^*(h-\phi)
+ c.c.]. \eea{r15} $L_{PF}$ is the Pauli-Fierz Lagrangian. The
mixing terms between $B_\mu$ and $h_{\mu\nu}, h, \phi$, which  do
not look like either a kinetic or a mass mixing can be easily
removed by adding first of all a gauge fixing term \beq L_{gf1}= a
\left|\partial_\nu h^{\mu\nu}-\frac{1}{2}\partial^\mu h + bB^\mu
\right|^2. \eeq{r16} With the judicious choice $a=-2$ and $b=m$ we
get rid of not only the mixing between $B_\mu$ and $h^{\mu\nu}$, but
also the kinetic mixing $\partial_\mu h^{*\mu\nu}\partial_\nu h$.
Notice that for this particular choice of parameters, $L_{gf1}$ does
not fix the scalar gauge transformation acting on $B_\mu,\phi$ and
given by eqs.~(\ref{r10}, \ref{r11}). This leaves room for adding a
second gauge fixing term, which may remove the remaining mixing
term, $m\partial^\mu B_\mu^*(h-3\phi)$. Indeed, this term can be
eliminated by a gauge fixing of the form \beq
L_{gf2}=c\,|\partial_\mu B^{\mu}+d(h - 3\phi)|^2, \eeq{r17} with
$c=-2$ and $d=m/2$. This fully fixes all gauge invariances.
Curiously, $L_{gf2}$ also kills the $|\partial_\mu B^\mu|^2$ term,
and the mass mixing between $h$ and $\phi$. We are finally left with
\bea L&=&h_{\mu\nu}^{*}(\Box-m^2)h^{\mu\nu} -
\frac{1}{2}h^*(\Box-m^2)h + 2B_\mu^*(\Box-m^2)B^\mu +
\frac{3}{2}\phi^{*}(\Box-m^2)\phi \nonumber \\
&&-\frac{1}{4}F_{\mu\nu}^2 + L_{int}. \eea{r18} Here all the kinetic
terms are diagonal, so that the propagators, all of which now have
the same pole\footnote{This is necessary to cancel spurious poles in
tree-level physical amplitudes.}, will behave nicely in the high
energy limit. It is worth noting that the ``wrong" sign for the
kinetic term of $h$ does not necessarily imply a propagating ghost.
In fact, such wrong signs usually appear when one performs a
covariant gauge fixing.

Now we turn our attention to the interaction terms. Schematically
\beq L_{int}=L_8+L_7+L_6+L_5+L_4, \eeq{r19} where $L_n$ contains the
operators having canonical dimension $n$, which are multiplied by a
factor $m^{4-n}$. For fixed $e$, in the high energy limit
$m\rightarrow 0$, the higher the $n$, the more potentially dangerous
the operator is. Notice that the gauge fixing terms
$L_{gf1},L_{gf2}$ are regular in the massless limit, so after the
minimal substitution $\partial_\mu \rightarrow D_\mu$ they only
generate a few harmless, power-counting renormalizable interactions.
A look at eq.~(\ref{r8}) reveals that by default each $\phi$ comes
with a factor $m^{-2}$, each $B_\mu$ with an $m^{-1}$, and each
$h_{\mu\nu}$ or $h$ with an $m^0$. Since the Pauli-Fierz Lagrangian
is quadratic in $h_{\mu\nu}$, we can at most have dimension-$8$
operators, which will necessarily involve a $\phi$ and a $\phi^*$.
Next we can have dimension-$7$ operators containing a $\phi$ and a
$B_\mu^*$, and so on. To pursue our analysis we must explicitly work
out these terms, taking good care of appropriate factors and signs.
After a tedious but straightforward calculation one finds \bea L_8
&=& \frac{ie}{m^4}\partial_\rho F^{\mu\rho} D_\mu D_\nu \phi^*
D^\nu\phi -
\frac{e^2}{4m^4}[5F_{\rho\sigma}^2\eta_{\mu\nu}+2F_{\mu\rho}
F^\rho_{\;\;\nu}]D^\mu\phi^* D^\nu\phi \nonumber\\
& &+ \frac{e^2}{4m^4}[(\partial^\mu F_{\mu\nu})^2-2(\partial_\mu
\partial_\rho F^{\mu\nu})F_\nu^{\;\;\rho}]\phi^*\phi-\,
\frac{e^2}{8m^4}(\partial^\rho F_{\mu\nu}^2)[\phi^*D_\rho \phi+\phi
D_\rho \phi^*], \label{r20}\\
L_7 &=& -\left\{\frac{ie}{m^3}D_\mu
B_\nu^*[F^{\mu\nu}\Box\phi+2\partial_\rho
F^{\mu\rho}D_\nu\phi]+ c.c.\right\}+\mathcal{O}(e^2),\label{r21} \\
L_6 &=& -\left\{\frac{2ie}{m^2}F^{\mu\nu}[\Box B_\mu^*B_\nu - 2D_\mu
B_\nu^*D^\rho B_\rho + 2D_\mu B_\rho^*D^\rho
B_\nu\,]+ c.c.\right\}\nonumber \\
& &+ \left\{\frac{ie}{m^2}F^{\mu\nu}[\partial^\rho
h_{\mu\rho}^*D_\nu\phi + 2h_{\mu\rho}^*D^\rho
D_\nu\phi]+c.c.\right\}+\mathcal{O}(e^2). \eea{r22}

Let us consider now a scaling limit: $m\rightarrow0$ and
$e\rightarrow0$, such that $e/m^4$=constant. In this limit the
Lagrangian simplifies enormously, becoming \beq L=L_{kin}
+\left(\frac{e}{m^4}\right)(\partial_\rho
F^{\mu\rho})\left[\frac{i}{2}\partial_\mu
\partial_\nu \phi^* \partial^\nu\phi+c.c. \right].
\eeq{r23} The above equation describes an effective field theory,
valid up to a finite cutoff \beq
\Lambda_4=\left(\frac{m^4}{e}\right)^{1/4}. \eeq{r24} It is the
spin-0 St\"uckelberg (a.k.a. Goldstone) boson that becomes strongly
coupled at high energies. This example illustrates the power of the
St\"{u}ckelberg formalism: it focuses precisely on the gauge modes
that give rise to the strong coupling. In the unitary gauge these
degrees of freedom are obscure, since they manifest as zero modes of
the free kinetic operator, and hence strong coupling phenomena
cannot be clarified so easily.

Notice that all the terms proportional to $e/m^4$ and $e/m^3$ in
eqs.~(\ref{r20}, \ref{r21}) are proportional to the equations of
motion. Thus one can eliminate them by appropriate {\em local} field
redefinitions. Of course, one will then introduce terms proportional
to $e^2/m^8$ and $e^2/m^6$. But one can hope that different possible
terms in the theory somehow conspire to cancel all the
$\mathcal{O}(e^2)$-terms\footnote{This may happen in particular if
other interactions exist, that are linear in $F_{\mu\nu}$ and mix
the spin 2 field with other more massive degrees of freedom. By
integrating out these additional degrees of freedom, one ends up
with additional EM interactions at $\mathcal{O}(e^2)$, that involve
only the spin 2 field.}. In this case the only dangerous
interactions are those linear in $e$ and {\em not} proportional to
the free equations of motion. If this is so, the Lagrangian reduces
to \beq L=L_{kin}+ \left\{\frac{ie}{m^2}F^{\mu\nu}[\partial^\rho
h_{\mu\rho}^*\partial_\nu\phi + 2h_{\mu\rho}^*\partial^\rho
\partial_\nu\phi+ 4\partial_\mu B_\nu^*\partial^\rho B_\rho - 4\partial_\mu
B_\rho^*\partial^\rho B_\nu]+\,c.c.\right\}, \eeq{r25} so that one
obtains a parametrically higher cutoff (in the regime $e\ll 1$,
which is the only one where our perturbative procedure makes sense):
\beq \Lambda_2=\frac{m}{\sqrt{e}}. \eeq{r26} Since the dangerous
interaction terms are linear in $e$, they can be eliminated neither
by a perturbative, local field redefinition nor by introducing
additional massive degrees of freedom.

\section{Adding Non-Minimal Terms}
In the most pessimistic scenario the cutoff of our effective field
theory is given by $(m^4/e)^{1/4}$. However as we will see now, this
cutoff can be pushed to the parametrically higher value of
$(m^3/e)^{1/3}$, without the help of extra massive degrees of
freedom, by adding appropriate non-minimal terms. The simplest
possibility is to add a dipole term. \bea L_{dipole} &=& ie\alpha
F^{\mu \nu} h^*_{\mu \rho}
h^\rho_{\;\;\nu} + c.c. \label{r27a} \\
&\rightarrow& ie\alpha F^\mu_{\;\;\nu} \left\{h^*_{\mu \rho} +
\frac{1}{m}(D_{\mu}B^*_{\rho}+D_{\rho}B^*_{\mu})
-\frac{1}{2m^2}(D_{\mu}D_\rho+D_{\rho}D_\mu)
\phi^*\right\}\times \nonumber \\
&&\left\{h^{\rho\nu } +
\frac{1}{m}(D^{\rho}B^{\nu}+D^{\nu}B^{\rho})-\frac{1}{2m^2}(D^{\rho}D^\nu+D^\nu
D^{\rho}) \phi\right\}+c.c. \,. \eea{r27} This non-minimal term will
again give rise to operators of various dimensions. In particular
the dimension-8 operator reads \beq L_{dipole}^8 =
-\frac{2ie\alpha}{m^4}(\partial_\rho F^{\mu\rho}) D_\mu D_\nu \phi^*
D^\nu\phi+\frac{e^2\alpha}{m^4}[F_{\rho\sigma}^2\eta_{\mu\nu}D^\mu\phi^*
D^\nu\phi+2F_{\mu\rho} F^\rho_{\;\;\nu}\phi^*D^\mu D^\nu\phi].
\eeq{r28} It is not surprising that the same operator $\partial_\rho
F^{\mu\rho} D_\mu D_\nu \phi^* D^\nu\phi$ shows up in the
$\mathcal{O}(e)$-terms in both eq.~(\ref{r20}) and~(\ref{r28}).
 In fact, antisymmetry of $F_{\mu\nu}$ allows only this
operator to appear. We can exploit this fact to choose $\alpha$ so
that the non-minimal Lagrangian does not contain any terms
proportional to $e/m^4$. This corresponds to choosing $\alpha=1/2$.

On the other hand, now we also have dimension-7 operators \beq
L_{dipole}^7 = -\left\{\frac{2ie\alpha}{m^3}F^{\mu\nu} (D_\mu
B_\rho^*+D_\rho B_\mu^*)D^\rho D_\nu \phi +
c.c.\right\}+\mathcal{O}(e^2). \eeq{r29} It is easy to see that no
choice of $\alpha$ can cancel all dimension-7 terms in the minimal
Lagrangian; not even to $\mathcal{O}(e)$. To make it even worse, the
non-minimal terms introduce new dimension-7 operators.

Thus the best we can do by adding the dipole term is to eliminate
terms proportional to $e/m^4$. In such a case, we take the
$m\rightarrow0$, $e\rightarrow0$ limit, keeping $e/m^3$=constant.
The non-minimal Lagrangian thereby reduces to \beq L=L_{kin}
-\left\{\frac{ie}{m^3}[(\partial_\rho F^{\mu\rho})(\partial_\mu
B_\nu^*\partial^\nu\phi+B_\mu^*\Box\phi) + (\partial_\rho
F^{\mu\nu})B_\mu^*\partial_\rho\partial_\nu\phi]+ c.c.\right\}.
\eeq{r30} Now both the longitudinal modes $\phi$ and $B_\mu$
participate in the strong coupling dynamics. But the cutoff is
parametrically higher than that of the minimal theory: \beq
\Lambda_3=\left(\frac{m^3}{e}\right)^{1/3}\gg\Lambda_4. \eeq{r31}

One may be tempted by the success of the above procedure to add
other non-minimal terms to further raise the cutoff scale. But that
does not help much, since all other possible non-minimal terms
contain at least dimension-6 operators to begin with (6 is a
quadrupole term). After the St\"{u}ckelberg procedure, addition of,
say, a quadrupole term produces operators up to dimension 10.
Although we can cancel the dimension-10 operators by a clever linear
combination of the possible quadrupole terms, we cannot get rid of
the dimension-9 operators. Thus not only that we gain nothing, but
actually we lower the UV cutoff of the theory. The conclusion,
therefore, is that $\Lambda_3$ is the highest we can raise the
cutoff to without adding additional massive degrees of freedom.
Notice that this cutoff is still lower than the ``optimistic'' one,
$\Lambda_2=m/\sqrt{e}$.

\section{Superluminality and Absence Thereof}
Many years ago Velo and Zwanziger~\cite{vz,v} discovered that
charged, massive fields of spin higher than one exhibit pathological
behavior in external (constant) electromagnetic fields. The
high-spin field may have modes that propagate faster than light, or
the number of propagating degrees of freedom may be different than
that of the free theory, or the Cauchy problem may become ill-posed.
All these pathologies are due to the fact that the free kinetic term
of high-spin fields exhibit gauge invariances, so that it has zero
modes. In the presence of electromagnetic interactions, these modes
acquire a non-vanishing but
non-canonical kinetic term, which may allow for some of
them to propagate superluminally, or which may not even be
hyperbolic. The formalism we employed in this paper is tailored to
single out the dynamics of precisely these zero modes, which are
none other than the St\"uckelberg fields. So, our formalism should
allow us to recover the results of Velo and Zwanziger, generalize
them, and simplify their derivation.

A first simplification is achieved by taking a convenient scaling
limit: \beq e\rightarrow 0,\qquad m\rightarrow 0, \qquad
\frac{eF_{\mu\nu}}{m^2}=\mbox{constant}. \eeq{m8} We further notice
that in a constant external electromagnetic background the
interaction terms $L_7$ and $L_6$ [eqs.~(\ref{r21}, \ref{r22})]
either vanish or become proportional to the free equations of motion
of the $B_\mu$ field. We can thus limit our analysis to solutions
where $B_\mu$ propagates on the light cone, so that only the scalar
$\phi$ exhibits  non-standard dynamics. By doing so, we may miss
some non-standard solutions in which the vector field $B_\mu$
propagates outside the light cone. So the following analysis will be
able to exhibit acausality and other defects of the spin 2 system,
but not to exclude them completely (they may disappear for the
scalar sector but reappear in the vector-scalar system).

Keeping this caveat in mind, we notice first that after taking the limit~(\ref{m8}) and
setting the $B_\mu$ field on shell, the only relevant interaction
terms in our non-minimal Lagrangian,
for a generic dipole coefficient $\alpha$, are
\beq
L_\phi=
-\frac{e^2}{4m^4}[(5-4\alpha)F_{\rho\sigma}F^{\rho\sigma}\eta^{\mu\nu}+
(2+8\alpha)F^{\mu\rho}
F_\rho^{\;\;\nu}]\partial_\mu\phi^* \partial_\nu\phi.
\eeq{r32}
These terms carry one derivative of $\phi$,
and another of $\phi^*$, so that in a constant electromagnetic
background they behave like additional kinetic terms for
the field $\phi$. This can potentially give rise to superluminal
propagation. As long as its kinetic term is concerned, $\phi$ will
experience a new effective background metric, different from
Minkowski:
\beq
\tilde{\eta}^{\mu\nu}= \left(\frac{3}{2}+\beta
F_{\rho\sigma}F^{\rho\sigma}\right)\eta^{\mu\nu}+\gamma F^{\mu\rho}
F_\rho^{\;\;\nu},
\eeq{r33}
where we have defined
\beq
\beta\equiv\frac{e^2}{4m^4}(5-4\alpha), \qquad
\gamma\equiv\frac{e^2}{4m^4}(2+8\alpha).
\eeq{r34}
Notice that eq.~(\ref{r33}) gives the {\em contravariant} metric.
For $\gamma=0$ the background metric $\tilde{\eta}^{\mu\nu}$ is
proportional to the Minkowski metric, therefore $\phi$ does not propagate
superluminally. This
case corresponds to a non-minimal Lagrangian with $\alpha=-1/4$. Notice also
that
even for this value of $\alpha$ the system ceases to be hyperbolic in a strong
field; precisely, when $e^2 F_{\mu\nu}F^{\mu\nu} /m^4=-1$.

However, things can be very different for other values of $\alpha$,
for example when $\alpha=0$, that corresponds to the minimal Lagrangian. Let us
consider two special cases of constant electromagnetic background: a
constant magnetic field, and a constant electric field.
\subsubsection*{Constant Magnetic Field: $\vec{B}=\hat{k}B$}
In this case the background metric~(\ref{r33}) reduces to
\beq
2\tilde{\eta}^{\mu\nu}=\text{diag}[-3-4\beta
B^2,\;3+(4\beta-2\gamma)B^2,\;3+(4\beta-2\gamma)B^2,\;3+4\beta B^2].
\eeq{r35}
On a plane perpendicular to $\vec{B}$, the wave speed is
different from $c$ in general; it is given by
\beq
\frac{v^2}{c^2}=1-\frac{2\gamma B^2}{4\beta
B^2+3}=1-\frac{e^2(1+4\alpha)B^2}{e^2(5-4\alpha)B^2+3m^4}.
\eeq{r36}
For different values of the parameter $\alpha$ one finds the
following results
\begin{eqnarray}
\begin{array}{cc}
  \underline{Value~of~\alpha} & \underline{Result} \\\\
  \alpha\geq 5/4 & $Superluminal propagation for$~
B^2>\frac{3m^4/e^2}{4\alpha-5} .\\\\
  1/2<\alpha<5/4 & $Loss of hyperbolicity for finite$~B.\\\\
  -1/4<\alpha\leq1/2 & $Subluminal propagation for any$~B.\\\\
  \alpha=-1/4 & v=c .\\\\
  \alpha<-1/4 & $Superluminal propagation for any$~B.\nonumber
\end{array}
\end{eqnarray}

We see that the choice $\alpha<-1/4$ is downright pathological,
since it gives superluminal propagation even for infinitesimally
small $B$. On the other hand, other values of $\alpha$ are good, at
least for sufficiently small values of $B$. In particular, for
$\alpha=-1/4$ the wave speed is $c$, as expected. The
superluminality reported by Velo and Zwanziger~\cite{vz,v}
corresponds to the case $\alpha\geq5/4$. In fact, one obtains the
same threshold value of $2m^2/3e$ by setting $\alpha-3=-1/16$.

At this point one may argue that both our minimal Lagrangian
($\alpha=0$), and the improved one with higher cutoff ($\alpha=1/2$)
are free of pathologies, because subluminal propagation by itself is
harmless. But before we draw a conclusion we need to consider other
cases.
\subsubsection*{Constant Electric Field: $\vec{E}=\hat{k}E$}
Here the background metric~(\ref{r33}) reduces to
\beq
2\tilde{\eta}^{\mu\nu}=\text{diag}[-3+(4\beta-2\gamma)E^2,
\; 3-4\beta E^2, \; 3-4\beta E^2, \; 3-(4\beta-2\gamma)E^2].
\eeq{r37}
Again, the wave speed on a plane perpendicular to $\vec{E}$
may not be $c$. Namely:
\beq
\frac{v^2}{c^2}=1+\frac{2\gamma E^2}{(4\beta-2\gamma)
E^2-3}=1+\frac{e^2(1+4\alpha)E^2}{e^2(4-8\alpha)E^2-3m^4}.
\eeq{r38}
Now we have the following results:
\begin{eqnarray}
\begin{array}{cc}
  \underline{Value~of~\alpha} & \underline{Result} \\\\
  \alpha\geq5/4 & $Subluminal propagation for any$~E. \\\\
  \alpha<5/4 & $Loss of hyperbolicity for finite$~E.\\\\
  -1/4<\alpha<1/2 & $Superluminal propagation for$ ~E^2
\geq \frac{3m^4/e^2}{4-8\alpha}. \\\\
  \alpha=-1/4 & v=c .\\\\
  \alpha<-1/4 & $Superluminal propagation for$
  ~ 0<E^2<\frac{3m^4/e^2}{4-8\alpha}. \nonumber
\end{array}
\end{eqnarray}
In this case too the domain $\alpha\geq-1/4$ is safe, at least for
sufficiently weak external fields. This is probably all we should
ask from our effective theory. To trust it for external backgrounds
$e F/m^2 \sim 1$ would require us to make the unreasonably strong
assumption that we can safely neglect all nonlinear, higher order
corrections in the external electromagnetic field. It is worth
noticing that even non-minimally coupled massive spin 3/2 fields
always mainfest  inconsistencies for large external
fields~\cite{dpw}, although they can be safe for small fields;
therefore, they too can make sense at most as effective low-energy
descriptions of some more fundamental theories. A particular
instance of non-minimally coupled charged spin 1 field theory,
arising from open strings, was studied in~\cite{an}. The same
phenomenon arises there as well: strong fields are
pathological,\footnote{And unreliable, since open strings in
external electric fields are unstable due to Schwinger pair
production at $|E|\sim M_{string}^2/e \leq m^2/e$~\cite{bp}.} but
weak  fields are not.
\subsubsection*{Generic Constant EM  Field}
For a generic {\em constant} electromagnetic background the same
conclusions hold. Indeed, whenever $\vec{E}\cdot \vec{B}=0$ and
$|\vec{E}|\neq |\vec{B}|$, there exists a frame in which the field
is either purely electric or purely magnetic, so that our previous
analysis applies trivially.

When $\vec{E}\cdot \vec{B}\neq 0$, there exists a frame in which
$\vec{E}$ is parallel to $\vec{B}$. Then, an analysis similar to our
previous cases tells us that in the range $\alpha\in[-1/4,1/2]$
superluminal propagation exists only in strong external fields;
namely: \beq \vec{E}^2 -\vec{B}^2 \geq \frac{3}{4\beta} +
\frac{\gamma}{ 2\beta} \vec{E}^2. \eeq{m9} This bound is weakest at
$\vec{B}=0$.

Finally, when the fields are perpendicular and equal in norm,
$\vec{E}\cdot \vec{B}=0$, $|\vec{E}|= |\vec{B}|\equiv E$, one can
choose a frame where the contravariant metric~(\ref{r33}) becomes
two-by-two block-diagonal
\bea \tilde{\eta} &\sim&
\left(\begin{array}{ll} A & 0 \\ 0 & I \end{array} \right), \label{m10} \\
A &=& \left(\begin{array}{ll} -1 -\frac{2}{3}\gamma E^2 &
-\frac{2}{3}\gamma E^2 \\
-\frac{2}{3}\gamma E^2 & 1-\frac{2}{3}\gamma E^2 \end{array}
\right). \eea{m11} In this metric background signals propagate with
three characteristic speeds: $c/\sqrt{1+\frac{2}{3}\gamma
E^2}$ or $c(\frac{2}{3}\gamma E^2 \pm
1)/(1+\frac{2}{3}\gamma E^2)$. Thus superluminality only appears for
$\gamma <0$, i.e., $\alpha < -1/4$.

\section{Conclusion}
This paper illustrates the power of the St\"uckelberg method for
understanding the dynamics of interacting high-spin fields. The
method renders free massive theories invariant under the same gauge
symmetries of massless theories by introducing redundant degrees of
freedom. They can be eliminated by using the gauge invariance,
thereby recovering the usual~\cite{sh} free Lagrangians. The
St\"uckelberg fields also make the {\em interacting} theory gauge
invariant. By appropriate covariant gauge fixings, field
redefinitions and scaling limits, one is then able to extract from
the full theory the sub-sector responsible for all pathologies of
the theory: strong coupling at finite energy scale, acausal
propagation in external fields etc. The St\"uckelberg formalism
unifies the description of all these phenomena.

We illustrated our point by analyzing only one of the simplest
high-spin systems: a spin 2 field coupled to electromagnetism. We
found out that the system possesses an intrinsic UV cutoff, no
higher than $\Lambda_2=m/\sqrt{e}$, and that its scalar
Goldstone/St\"uckelberg sector is where the acausal behavior found by
Velo and Zwanziger manifests itself. Generalizations of this example
are under way~\cite{pr}. They are both intriguing and subtle, since
starting from spin 3, the St\"uckelberg sector itself contains spin
2 fields which, in particular, does not admit a smooth massless
limit~\cite{vvdz}. What happens then when we try to take a scaling
limit where $m \rightarrow 0$? Irrespective of the answer to this
question, the behavior of high-spin fermions should not differ
substantially  from that of bosons, since the source of
strong-coupling pathologies is the existence of gauge invariances in
the massless, free kinetic term, independently of their statistics
(see e.g. refs~\cite{fpt,d2,vz,v,dpw}).

Gravitational interactions of high-spin fields are also of paramount
importance, since they
are truly universal. If an intrinsic cutoff is found for such interactions,
it signals the
ultimate limit of any local effective field theory description of interacting
high-spin massive fields.

\subsection*{Acknowledgments}
MP is supported in part by NSF grant PHY-0245068 and by a Marie Curie
Excellence Chair, contract MEXC-CT-2003-509748 (SAG@SNS).

\end{document}